\documentclass[aps,prb,twocolumn,showpacs]{revtex4}

\usepackage{hyperref}
\usepackage{graphicx}
\usepackage{epsfig,epsf,bm}

\begin{document}
\title{Ground state of graphite ribbons with zigzag edges}
\author{Yu-Li Lee}
\email{c00lyl00@nchc.gov.tw} \affiliation{Division of Nanoscience,
National Center for High-performance Computing, Hsinchu, Taiwan,
R.O.C.}

\author{Yu-Wen Lee}
\email{ywlee@mail.thu.edu.tw} \affiliation{Physics Department,
Tunghai University, Taichung, Taiwan, R.O.C.}
\begin{abstract}
 We study the interaction effects on the ground state of nanographite
 ribbons with zigzag edges. Within the mean-field approximation, we
 found that there are two possible phases: the singlet superconducting (SS)
 phase and the excitonic insulator (EI). The two phases are
 separated by a first-order transition point. After taking into
 account the low-lying fluctuations around the mean-field solutions,
 the SS phase becomes a spin liquid phase with one gapless charge mode.
 On the other hand, all excitations in the EI phase, especially the
 spin excitations, are gapped.
\end{abstract}
\maketitle

\section{Introduction}

After the discovery of low-dimensional materials such as
fullerenes and carbon nanotubes, the research on the $sp^2$
network systems has been attracting much attention. The
nanographite ribbon is one of the most simple and fundamental
fragments of the $sp^2$ network and represents a new class of
mesoscopic systems. In this system, the boundary regions play an
important role so that the edge effects may influence strongly the
$\pi$-electron states near the Fermi surface.

\begin{figure}
\includegraphics[width=0.75\columnwidth]{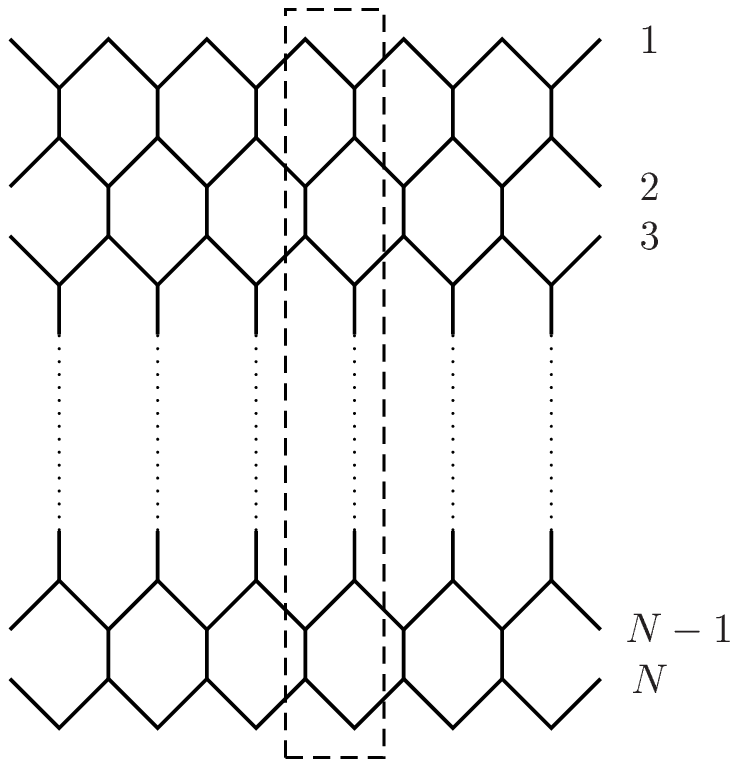}
\caption{The structure of graphite ribbons with zigzag edges. The
carbon atoms are located at the corners of each hexagons. The
rectangle with the dashed line is the unit cell.} \label{unit}
\end{figure}

There are two basic shapes of regular graphite edges -- zigzag and
armchair edges (see Fig. \ref{unit}). The study of electronic
states of hydrogen-terminated graphite ribbons reveals that the
ribbons with zigzag edges possess partly flat bands at the Fermi
level, which correspond to the electronic states localized in the
near vicinity of the edges\cite{NFDD,WFAS,MNF}. Specially, the
highest valence band and the lowest conduction band are always
degenerate at $ka_0=\pi$ with the lattice spacing $a_0\approx
2.46$ \AA. (Hereafter, we will set $a_0=1$.) The localized edge
states are of special interest because of their relatively large
contribution to the density of states (DOS) at the Fermi surface,
which results in the Curie-like temperature dependence of the
Pauli susceptibility\cite{WFAS} and zero-conductance resonances in
the nanographite ribbon junctions\cite{WS}. It was reported that
the zigzag ribbons do not undergo bond alternations along the
ribbon axis for reasonable strength of electron-phonon
interactions because of the non-bonding character of the edge
states\cite{WSF}. In other words, the partly flat bands are stable
against the Peierls instability. In addition, the flat edge states
exist not only in the single-layered zigzag ribbons but also in
the stacked layers of zigzag ribbons in a manner of the $AB$
stacking\cite{MNF}, in which half of the carbon atoms of one
ribbon are located directly above the center of each hexagon on
the neighboring ribbons.

In this paper, we are interested in the low-energy physics of the
graphite ribbons with zigzag edges. In that case, it suffices to
consider the edge states only. A simple power-counting indicates
that all four-fermion interactions are relevant operators around
the free-fermion fixed point (see section III). Therefore, we
expect that the electron-electron interactions will change the
free-fermion picture drastically. Based on the renormalization
group (RG) analysis, we propose a model Hamiltonian with an O($4$)
symmetry to describe the low energy physics of this system. A
mean-field treatment of this Hamiltonian results in two possible
zero temperature phases -- SS and EI phases separated by a
first-order transition point. For the EI phase\cite{HR}, the
singlet and triplet excitonic order parameters can coexist due to
the underlying O($4$) symmetry\cite{Bal}. As a result, there is a
degeneracy for the onset of the charge density wave (CDW) and spin
density wave (SDW) ordering within the mean-field approximation.
However, in one dimension, the low-lying fluctuations around the
mean-field solutions are so strong that the long range orders
obtained by the mean-field theory are destroyed and the true
ground states exhibit the algebraic (SS) or short-ranged (EI)
orders. Consequently, the SS phase becomes a spin liquid with one
gapless charge mode, whereas the EI phase turns into an insulating
phase in which all excitations are gapped and the broken O($4$)
symmetry is restored.

The rest of the paper is organized as follows: In section II, we
give a description of the system and discuss the action describing
the dynamics of the flat edge states. We present the
renormalization group (RG) analysis of the action in section III
and propose an effective Hamiltonian which describes the low
energy physics. The mean-field theory and the derivation of
low-energy effective actions are given in section IV and V,
respectively. The last section is devoted to the discussions and
conclusions.

%%%%%%%%%%%%%%%%%%%%%%%%%%%%%%%%%%%%%%%%%%%%%%%%%%%%%%%%%%%%%%%%%%%%%%%%%%%%
\section{The model system}

\begin{figure}
\includegraphics[width=0.7\columnwidth]{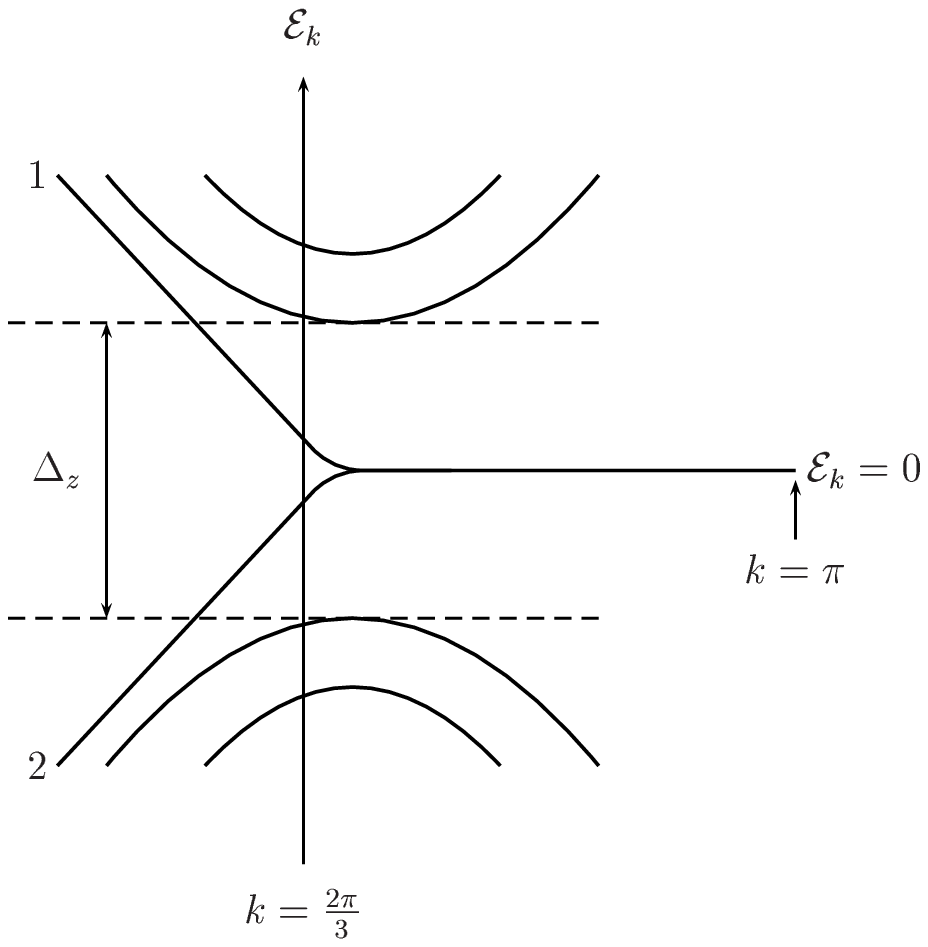}
\caption{The schematic figure of the energy band dispersion near
$E=0$. $\Delta_z$ is given by Eq. (\ref{band2}). The band indices
$1$ and $2$ indicate the two lowest bands.} \label{disp}
\end{figure}

The structure of the graphite ribbon with zigzag edges and the
schematic figure of its energy band dispersion near $E=0$ are
shown in Figs. \ref{unit} and \ref{disp}, respectively. The almost
flat bands appear within the region $\frac{2\pi}{3}\leq |k|\leq
\pi$. For an H-terminated single-layered zigzag ribbon, the
dispersion relations of the two lowest bands close to $k=\pi$
obtained from the tight-binding model has the approximate
form\cite{NFDD,WFAS}
\begin{equation}
 {\cal E}_{1(2)k}=\pm 2tND_k^{N-1}\left(1-\frac{D_k}{2}\right) \ ,
           \label{band1}
\end{equation}
where $D_k=2\cos{(\frac{k}{2})}$, $t$ is the hopping matrix
element and $N$ is the number of zigzag lines (see Fig.
\ref{unit}). Eq. (\ref{band1}) indicates that the two lowest bands
are degenerate at the Fermi point $k=k_0=\pi$. Around that point,
${\cal E}_{1(2)k}\approx \pm 2Nt|k-k_0|^{N-1}$. The energy gap to
the next band $\Delta_z$ as shown in Fig. \ref{disp} is given
by\cite{NFDD,WFAS}
\begin{equation}
 \Delta_z=4t\cos{\left[\frac{(N-1)\pi}{2N+1}\right]} \ . \label{band2}
\end{equation}
Since the energy scale we are considering is much lower than
$\Delta_z$, only these two lowest bands are involved and the
electron operator can be expanded around the Fermi point $k_0$
\begin{equation}
 \Psi_{\alpha}(\vec{x})\approx e^{ik_0x}\left[\psi_{1\alpha}(x)
      u_1(y)+\epsilon_{\alpha \beta}\psi_{2\beta}^{\dagger}(x)
      u_2(y)\right] \ , \label{fermi1}
\end{equation}
such that $\psi_{a\alpha}|0\rangle =0$, where $a=1,2$ is the band
index, and $\alpha =\pm 1$ corresponds to spin up and down,
respectively. (We assume that the $x$-axis is parallel to the
ribbon axis.) The $\psi$-fermions describe the low-energy degrees
of freedom. $u_{1,2}(y)$ are real functions which satisfy the
orthonormal condition: $\int dy \ u_a(y)u_b(y)=\delta_{ab}$. Under
the operation of reflection about the line along the ribbon axis
at the middle of the ribbon, denoted by a unitary operator ${\cal
R}$, we have ${\cal R}u_1=u_2$. (Note that this results in
$u_1^2=u_2^2$.)

Up to the four-fermion interactions, the most general form of the
action describing the dynamics of $\psi$-fermions in the
imaginary-time formulation is given by
\begin{eqnarray*}
 I=I_0+\int d\tau dx \ L_1 \ ,
\end{eqnarray*}
where
\begin{eqnarray}
 I_0 &=& \int \frac{d\omega}{2\pi}\int^{\Lambda}_{-\Lambda}\frac{dk}
     {2\pi}~\tilde{\psi}^{\dagger}_{a\alpha}\left(-i\omega +v|k|^m
     \right)\tilde{\psi}_{a\alpha} \ , \label{act1} \\
 L_1 &=& \frac{g_1}{2}\left(\rho_1^2+\rho_2^2\right)+g_2\rho_1\rho_2
     +4g_3\bm{J}_1\cdot \bm{J}_2 \nonumber \\
     & & +g_4(N_1N_2+{\rm H.c.}) \ . \label{act2}
\end{eqnarray}
Here $\tilde{\psi}_{a\alpha}(k)\equiv \int dx \
e^{-ikx}\psi_{a\alpha}(x)$, $m>1$ ($m=N-1$ for the single-layered
graphite ribbons), $v=2Nt$, $\Lambda$ is an UV cut-off for the
momentum, and
\begin{eqnarray}
 \rho_a &=& \psi_{a\alpha}^{\dagger}\psi_{a\alpha} \ , \label{op1} \\
 \bm{J}_a &=& \frac{1}{2}\psi_{a\alpha}^{\dagger}
        (\bm{\sigma})_{\alpha \beta}\psi_{a\beta} \ , \nonumber \\
 N_a &=& \frac{i}{2}\epsilon_{\alpha \beta}\psi_{a\alpha}\psi_{a\beta}
     \ . \nonumber
\end{eqnarray}
In Eq. (\ref{act1}), we have replaced $k-k_0$ by $k$. $\rho_a$,
$\bm{J}_a$, and $N_a$ defined above are, respectively, the charge
density, the spin density, and the singlet Cooper pair in the band
$a$. $g_1$ is the intra-band interaction, and the inter-band
interactions are described by the density-density interaction
$g_2$, the exchange interaction $g_3$, and the singlet Cooper-pair
tunneling $g_4$. (Note that the triplet Cooper-pair tunneling term
vanishes in the present case because of the Fermi statistics.) We
have to emphasize that, regardless of the graphite ribbons being
H-terminated, not H-terminated, single-layered, or
stacked-layered, Eqs. (\ref{act1}) and (\ref{act2}) describe the
dynamics of the flat edges states as long as they exist. The short
distance structure only affects the values of the parameters $g_i$
($i=1,\cdots ,4$), $v$, and $m$, which will ultimately determine
in which phase the system is located.

To obtain the values of the couplings $g_i$ ($i=1,\cdots ,4$), we
consider the short-ranged electron-electron interactions described
by the interacting Hamiltonian
\begin{widetext}
\begin{equation}
 {\rm H}_{int}=\frac{1}{2}\int d^2x_1d^2x_2 \ :\Psi^{\dagger}_{\alpha}
              (\vec{x}_1)\Psi^{\dagger}_{\beta}(\vec{x}_2)V(\vec{x}_1
              -\vec{x}_2)\Psi_{\beta}(\vec{x}_2)\Psi_{\alpha}
              (\vec{x}_1): \ . \label{inth1}
\end{equation}
By inserting Eq. (\ref{fermi1}) into Eq. (\ref{inth1}), we obtain
\begin{equation}
 g_1=U_1(0) \ , ~~ g_2=-U_1(0)+\frac{U_2(0)}{2} \ , ~~
 -2g_3=U_2(0)=g_4 \ , \label{coupl1}
\end{equation}
where $U_i(q)=\int dx \ e^{-iqx}V_i(x)$ ($i=1,2$), and
\begin{eqnarray*}
 V_1(x_1-x_2) &=& \int dy_1dy_2 \ u_1^2(y_1)V(\vec{x}_1-\vec{x}_2)
              u_1^2(y_2) \ , \\
 V_2(x_1-x_2) &=& \int dy_1dy_2 \ u_1(y_1)u_2(y_1)V(\vec{x}_1-\vec{x}_2)
              u_1(y_2)u_2(y_2) \ .
\end{eqnarray*}
\end{widetext}
Especially, for a Hubbard-like interaction
$V(\vec{x}_1-\vec{x}_2)=V_0\delta (\vec{x}_1-\vec{x}_2)$, we have
$g_1=U=-2g_2=-2g_3=g_4$ where $U=V_0\int dy~u_1^4(y)$. We see that
$g_1,g_4>0$ and $g_2,g_3<0$ when the interactions between
electrons are repulsive.

The symmetries of the action $I$ is the charge U($1$) (denoted by
U$_c$($1$)) where the $\psi$-fermions transform as:
$\psi_{1\alpha}\rightarrow e^{i\theta}\psi_{1\alpha}$ and
$\psi_{2\alpha}\rightarrow e^{-i\theta}\psi_{2\alpha}$, the spin
SU($2$), and the Z$_2$ (particle-hole) where the $\psi$-fermions
transform as: $\psi_{1\alpha}\leftrightarrow \psi_{2\alpha}$. In
this paper, we restrict ourselves to the undoped case. The
chemical potential term breaks the Z$_2$ symmetry and thus will
not be generated by renormalization.

%%%%%%%%%%%%%%%%%%%%%%%%%%%%%%%%%%%%%%%%%%%%%%%%%%%%%%%%%%%%%%%%%%%%%%%%%%%%%%
\section{Renormalization group analysis}

We first show that $L_1$ is a relevant perturbation to the
free-fermion action $I_0$. By integrating out the fast modes with
$\Lambda /s<|k|<\Lambda$, we find that $I_0$ is a fixed point of
the following RG transformation
\begin{equation}
 k\rightarrow k/s \ , ~~~~ \omega \rightarrow \omega s^{-m} \ ,
 ~~~~ \tilde{\psi}_{a\alpha}\rightarrow \xi \tilde{\psi}_{a\alpha} \ ,
      \label{rg1}
\end{equation}
with $\xi =s^{m-1/2}$. Under the RG transformation (\ref{rg1}),
the couplings in $L_1$ transform as $g_i^{\prime}=s^{m-1}g_i$
($i=1,\cdots,4$). We see that in one dimension all couplings are
relevant and have the same scaling dimension around the
free-fermion fixed point $I_0$. To tell which terms dominate the
low energy physics, a controllable approximation to organize the
quantum fluctuations is necessary. Here we adopt the
$\epsilon$-expansion\cite{erg}. That is, we extend the spatial
dimension from $1$ to $d$ and then use $\epsilon =d_u-d$ as the
expansion parameter to compute the RG functions, where $d_u=m$ is
the upper-critical dimension. For small $N$ (and thus small $m$),
the $\epsilon$-expansion may be reliable. For larger $N$, it is
hoped that there is no qualitative difference in physical
properties from small $N$ to large $N$ as long as $N$ is finite.

Next we would like to examine the quantum fluctuations up to the
one-loop order. Defining the dimensionless couplings:
$\lambda_i=\Lambda^{-\epsilon}cg_i/v$ ($i=1,\cdots ,4$) where
$c=S_{N-1}/[2(2\pi)^{N-1}]$ and $S_d$ is the area of a
$d$-dimensional sphere, and calculating the one-loop corrections
to the coupling constants $g_i^{\prime}$ ($i=1,\cdots ,4$), we
obtain the one-loop RG equations within the $\epsilon$-expansion:
\begin{eqnarray}
 \frac{d\lambda_1}{dl} &=& \epsilon \lambda_1-\lambda_1^2-
      \lambda_4^2 \ , \nonumber \\
 \frac{d\lambda_2}{dl} &=& \epsilon \lambda_2-\lambda_2^2-
      3\lambda_3^2-\lambda_4^2 \ , \nonumber \\
 \frac{d\lambda_3}{dl} &=& \epsilon \lambda_3-2\lambda_3(\lambda_2
      -\lambda_3) \ , \nonumber \\
 \frac{d\lambda_4}{dl} &=& \epsilon \lambda_4-2\lambda_4(\lambda_1
      +2\lambda_2) \ . \label{rg2}
\end{eqnarray}
Eq. (\ref{rg2}) is solved numerically up to a scale $l^*$, at
which point the largest coupling $\lambda_{max}={\rm
max}\{\lambda_i(l^*)\}$ satisfies $\tilde{U}\ll \lambda_{max}\ll
1$, where $\tilde{U}\equiv \Lambda^{-\epsilon}cU/v$. This allows
us to ignore the higher-order terms [$O(\lambda_i^3)$] in the RG
equations. As $\tilde{U}\rightarrow 0$, $l^*\rightarrow \infty$,
and we need only analyze the asymptotic large $l$ behaviors of Eq.
(\ref{rg2}). We find that for $\epsilon >0$, $\lambda_2$ and
$\lambda_4$ become divergent first while the value of $\lambda_1$
approaches zero and that of $\lambda_3$ remains small when
$\lambda_{2,4}=O(1)$.

Based on this analysis, we propose that the low-energy physics of
the graphite ribbons with zigzag edges is described by the
following model Hamiltonian:
\begin{eqnarray}
 {\rm H} &=& \int \frac{dk}{2\pi} \ \epsilon_k
         \tilde{\psi}_{a\alpha}^{\dagger}(k)\tilde{\psi}_{a\alpha}(k)
         \label{h1} \\
         & &+\int dx \ [g(N_1N_2+{\rm H.c.})-\tilde{g}\rho_1\rho_2] \ ,
         \nonumber
\end{eqnarray}
where $g,\tilde{g}>0$ and $\epsilon_k=v|k|^m$. The symmetry of
${\rm H}$ is U$_c$($1$)$\times$SU($2$)$\times$SU($2$)$\times$Z$_2$
where the SU($2$)$\times$SU($2$) symmetry corresponds to the
independent spin rotations in each band. The enlarged symmetry
(SU($2$)$\rightarrow$ SU($2$)$\times$SU($2$)) arises from the
suppression of the $g_3$ term under the RG transformation. The
values of $g$ and $\tilde{g}$ depend on $v$ and the initial values
of $g_i$'s ($i=1,\cdots 4$). In the following, we shall treat the
coupling constants $g$ and $\tilde{g}$ as free parameters and
study the possible phases of the Hamiltonian (\ref{h1}).

%%%%%%%%%%%%%%%%%%%%%%%%%%%%%%%%%%%%%%%%%%%%%%%%%%%%%%%%%%%%%%%%%%%%%%%%%%%%
\section{Mean-field theory}

The $g$ term in the Hamiltonian (\ref{h1}) describes the process
of singlet Cooper-pair tunneling between two bands and favors the
singlet superconductor when it becomes divergent, whereas a strong
$\tilde{g}$ term enhances the fluctuations of electron-hole pairs
or excitonic ordering between band $1$ and band $2$\cite{KM}. The
latter can be represented by the singlet or triplet excitonic
order parameters according to the angular momentum it carries.
Motivated by this observation, we define the following order
parameters:
\begin{eqnarray}
 \hat{O}_{sa}(x) &=& g\psi_{a\uparrow}(x)\psi_{a\downarrow}(x) \ ,
                 \label{op2} \\
 \hat{\Phi}_s(x) &=& i\frac{\tilde{g}}{2}\epsilon_{\alpha \beta}
                 \psi_{1\alpha}(x)\psi_{2\beta}(x) \ , \nonumber \\
 \hat{\bm{\Phi}}_t(x) &=& \frac{\tilde{g}}{2}\epsilon_{\alpha \lambda}
                 \psi_{2\lambda}(x)(\bm{\sigma})_{\alpha \beta}
                 \psi_{1\beta}(x) \ . \nonumber
\end{eqnarray}
With the help of the Hubbard-Stratonovich transformation, the
corresponding action of the Hamiltonian (\ref{h1}) in the
imaginary-time formulation can be written as
\begin{eqnarray}
 S &=& \int \frac{d\omega}{2\pi}\frac{dk}{2\pi} \
   \tilde{\psi}_{a\alpha}^{\dagger}(-i\omega +\epsilon_k)
   \tilde{\psi}_{a\alpha} \nonumber \\
   & & +\int d\tau dx \ ({\cal L}_1+{\cal L}_2+{\cal L}_3) \ ,
   \label{act3}
\end{eqnarray}
where
\begin{eqnarray*}
 {\cal L}_1 &=& \frac{1}{g}\hat{O}_{s1}\hat{O}_{s2}-\hat{O}_{s1}
            \psi_{2\uparrow}\psi_{2\downarrow}-\hat{O}_{s2}
            \psi_{1\uparrow}\psi_{1\downarrow}+{\rm H.c.} \ , \\
 {\cal L}_2 &=& \frac{2}{\tilde{g}}|\hat{\Phi}_s|^2-\left(
            i\hat{\Phi}_s^{\dagger}\epsilon_{\alpha \beta}
            \psi_{1\alpha}\psi_{2\beta}+{\rm H.c.}\right) \ , \\
 {\cal L}_3 &=& \frac{2}{\tilde{g}}|\hat{\bm{\Phi}}_t|^2-\left(
            \hat{\bm{\Phi}}_t^{\dagger}\cdot \epsilon_{\alpha \lambda}
            \psi_{2\lambda}(\bm{\sigma})_{\alpha \beta}\psi_{1\beta}+
            {\rm H.c.}\right) \ .
\end{eqnarray*}

The mean-field treatment of the action $S$ (\ref{act3}) is to
assume the presence of bosonic mean fields, neglect the
fluctuations of order parameters, and finally, integrate out the
fermionic degrees of freedom to derive the effective potential.
Now we consider the mean-field ansatz: $O_{sa}=\langle
\hat{O}_{sa}\rangle$, $\Phi_s=\langle \hat{\Phi}_s\rangle$, and
$\bm{\Phi}_t=\langle \hat{\bm{\Phi}}_t\rangle$. The effective
potential is given by
\begin{eqnarray}
 V &=& \frac{1}{g}(O_{s1}O_{s2}+{\rm C.c.})+\frac{2}{\tilde{g}}\left(
   |\Phi_s|^2+|\bm{\Phi}_t|^2\right) \nonumber \\
   & & -\int \frac{d\omega}{2\pi}\frac{dk}{2\pi} \ \ln{\left[\frac{
   F(\omega ,k)}{F_0(\omega ,k)}\right]} \ , \label{effv1}
\end{eqnarray}
where $F_0(\omega ,k)=(\omega^2+\epsilon_k^2)^2$, and
\begin{eqnarray*}
 F(\omega ,k) &=& (\omega^2+\epsilon_k^2)^2+\left|O_{s1}^*O_{s2}^*
              +\Delta^T\Delta \right|^2 \nonumber \\
              & & +(\omega^2+\epsilon_k^2)\left(|O_{s1}|^2+
              |O_{s2}|^2+2\Delta^{\dagger}\Delta \right) \ .
\end{eqnarray*}
Here $\Delta =(\Phi_s,\bm{\Phi}_t)^T$ is the excitonic order
parameter. In Eq. (\ref{effv1}), we have set $V=0$ for free
fermions.

The ground state is determined by the minimum of the effective
potential $V$ and the solutions of the mean-field equations are
given by
\begin{equation}
 O_{s1}=O_{s2}^*=\Delta_se^{i\phi_0} \ , ~~ \Delta=0 \ , \label{sol1}
\end{equation}
where $\Delta_s=C_mv^{-\frac{1}{m-1}}g^{\frac{m}{m-1}}$ with
$C_m=a_m^{\frac{m}{m-1}}$ and
$a_m=B\left(\frac{1}{2m},\frac{1}{2}-\frac{1}{2m}\right)/(4m\pi)$,
and
\begin{equation}
 \Delta =\Delta_0e^{i\theta_0} \ , ~~ O_{s1}=0=O_{s2} \ , \label{sol2}
\end{equation}
where $\Delta_0$ is a real vector with the fixed length:
$\Delta_e=\sqrt{\Delta_0^T\Delta_0}=C_mv^{-\frac{1}{m-1}}\tilde{g}^{\frac{m}{m-1}}$.
Here $B(x,y)$ is the beta function. The solutions (\ref{sol1}) and
(\ref{sol2}) correspond to the SS and EI phases, respectively.
Especially, in the EI phase, $\Phi_s\neq 0$ can coexist with
$\bm{\Phi}_t\neq 0$. It is well known\cite{KM,VKR} that the
singlet ($\Phi_s$) and triplet ($\bm{\Phi}_t$) excitonic orders
are accompanied by the appearance of CDW and SDW. In our case, the
onset of both ordering is degenerate on account of the O($4$) or
SU($2$)$\times$SU($2$) symmetry of the action ({\ref{act3}). The
weak interactions which are neglected in the Hamiltonian
(\ref{h1}) will lift this degeneracy. A residual Coulomb
interaction like the $g_3$ term favors the triplet excitonic
order, while the electron-phonon interactions favor the singlet
excitonic order\cite{VKR}. Note that naively the hidden symmetry
of the EI phase is U$_f$($1$)$\times$O($4$), where the U$_f$($1$)
symmetry corresponds to the transformation:
$\psi_{a\alpha}\rightarrow e^{i\tilde{\theta}}\psi_{a\alpha}$.
However, it is not a symmetry of the Hamiltonian (\ref{h1}) and we
will see later that after taking into account the fluctuations the
value of $\theta_0$ cannot be arbitrary.

The point $g=\tilde{g}$ corresponds to the first-order phase
transition because at that point the ground-state energies of the
SS and EI phases are equal and the corresponding order parameters
do not vanish. Thus, the values of the order parameters change
discontinuously from the SS to EI phases or vice versa. This is
consistent with the RG flow obtained from the
$\epsilon$-expansion. Since the RG equation (\ref{rg2}) does not
exhibit any IR stable fixed point, the phase transition cannot be
the second-order one. To sum up, the zero-temperature phase
diagram within the mean-field approximation is shown in Fig.
\ref{pd}.
\begin{figure}
\includegraphics[width=0.8\columnwidth]{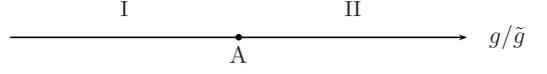}
\caption{The mean-field phase diagram. Region I is the EI phase
and region II is the SS phase. The point A ($g=\tilde{g}$) is a
first-order transition point} \label{pd}
\end{figure}

%%%%%%%%%%%%%%%%%%%%%%%%%%%%%%%%%%%%%%%%%%%%%%%%%%%%%%%%%%%%%%%%%%%%%%%%%%%%
\section{Low-energy effective action}

To examine the stability of the mean-field solutions obtained
above, we have to consider the fluctuations around the mean-field
ground state.

In the SS phase, $\hat{O}_{s1}$ and $\hat{O}_{s2}$ can be
parametrized as: $\hat{O}_{s1}=\Delta_se^{i\phi_1}$ and
$\hat{O}_{s2}=\Delta_se^{-i\phi_2}$. On the other hand, the
excitonic order parameter $\Delta$ becomes the ordinary bosonic
field. Inserting these into the action (\ref{act3}) and
integrating out the fermionic fields and $\Delta$, we obtain the
low-energy effective theory in the SS phase
\begin{eqnarray}
 {\cal L}_s &=& \frac{K_0}{2}\left[\frac{1}{v_s}(\partial_{\tau}\phi_+)^2
            +v_s(\partial_x\phi_+)^2\right] \label{lag1} \\
            & & +\frac{\bar{K}_0}{2}\left[\frac{1}{\bar{v}_s}
            (\partial_{\tau}\phi_-)^2+\bar{v}_s(\partial_x\phi_-)^2
            \right]+\mu_0\cos{\phi_-} \ , \nonumber
\end{eqnarray}
where $\phi_{\pm}=\phi_1\pm \phi_2$,
$K_0=\bar{K}_0=\frac{m-1}{2}\sqrt{a_mb_m}$,
$v_s=\bar{v}_s=m\sqrt{\frac{b_m}{a_m}}v^{\frac{1}{m}}\Delta_s^{1-\frac{1}{m}}$,
and $b_m\propto (\Delta_z/\Delta_s)^{1-\frac{1}{m}}\gg 1$ is a
non-universal constant. The $\phi_+$-field carries charge two and
spin zero while the quantum numbers of $\phi_-$ are $Q=0$ and
${\cal S}=0$. The cosine term in ${\cal L}_s$ opens a gap in the
$\phi_-$ sector because $\bar{K}_0>(8\pi)^{-1}$. The charge U($1$)
symmetry forbids the terms like $\cos{(\beta \phi_+)}$ or
$\sin{(\beta \phi_+)}$. Therefore, the $\phi_+$ sector remains
gapless. After taking into account the fluctuations, the
mean-field gap $\Delta_s$ becomes the spin gap and the SS phase
is, in fact, a spin liquid (SL) phase with one gapless charge
mode, which is similar to the $C1S0$ phase in the two-leg Hubbard
ladder\cite{BF}. Besides, the long range SS order turns into the
algebraic one due to the gapless $\phi_+$-field and the SS
fluctuations are enhanced compared with the free fermions.

In the EI phase, the excitonic order parameter $\Delta$ can be
written as: $\Delta =\Delta_e\Phi e^{i\theta}$ where $\Phi$ is a
real vector satisfying $\Phi^T\Phi=1$ and transforms with the
fundamental representation of O($4$), whereas $\hat{O}_{s1}$ and
$\hat{O}_{s2}$ are ordinary bosonic fields. The quantum numbers of
the $\theta$-field are $Q=0$ and ${\cal S}=0$ and the ones carried
by the $\Phi$ sector consist of $(Q,{\cal S})=(0,0),(0,1)$. The
low-energy effective theory in this case can be obtained by
inserting the above parametrization into the action (\ref{act3})
and integrating out the fermionic fields, $\hat{O}_{s1}$, and
$\hat{O}_{s2}$, and then we have
\begin{eqnarray*}
 S_{EI}=\int d\tau dx \ ({\cal L}_{\theta}+{\cal L}_{\Phi}) \ ,
\end{eqnarray*}
where
\begin{eqnarray}
 {\cal L}_{\theta} &=& \frac{K}{2}\left[\frac{1}{v_e}(\partial_{\tau}\theta)^2
                   +v_e(\partial_x\theta)^2\right]+\lambda \cos{(2\theta)} \ ,
                   \label{lag2} \\
 {\cal L}_{\Phi} &=& \frac{\rho_o}{2}\left[\frac{1}{v_o}\left(
                 \partial_{\tau}\Phi\right)^2+v_o\left(\partial_x\Phi \right)^2
                 \right] \ , \label{lag3}
\end{eqnarray}
Here $K=\rho_o=2(m-1)\sqrt{a_m\bar{b}_m}$,
$v_e=v_o=m\sqrt{\frac{\bar{b}_m}{a_m}}v^{\frac{1}{m}}\Delta_e^{1-\frac{1}{m}}$,
and $\bar{b}_m\propto (\Delta_z/\Delta_e)^{1-\frac{1}{m}}\gg 1$ is
a non-universal constant. In Eq. (\ref{lag2}), we keep the leading
cosine term only. Because $K>(2\pi)^{-1}$, the cosine term in
${\cal L}_{\theta}$ is a relevant operator and the $\theta$ sector
acquires a gap. Beisdes, $\langle \theta \rangle$ is pinned at
some value $\theta_0$ which depends on the short distance physics.
The dynamics of the $\Phi$ sector is described by the O($4$)
non-linear $\sigma$ model. In general, there can be two sources to
change the low-energy behavior of the non-linear $\sigma$ model.
The first one is the existence of a term linear in
$\partial_{\tau}\Phi$, which follows from the analysis of the
equations of motion (EOM) and Ward identities\cite{Lut}. As
discussed in Ref. \onlinecite{Lut}, the existence of this term
relies on the nonvanishing expectation values of the corresponding
conserved charges, i.e. $\langle \bm{J}_a \rangle \neq 0$ in the
present case. But $\langle \bm{J}_a \rangle \neq 0$ implies the
long range ferromagnetic (FM) order and it is forbidden in $1+1$
dimensions by the Mermin-Wagner-Coleman theorem\cite{foot1}. The
second one is the appearance of the $\theta$-term, which results
from the topological consideration and has no effects on the EOM.
In our case, ${\cal L}_{\Phi}$ does not contain such a term
because the homotopy group $\Pi_2[{\rm O}(4)]=0$. As a result, in
the long wavelength limit, ${\cal L}_{\Phi}$ starts from the
second derivatives of space and time as shown in Eq. (\ref{lag3}).
Therefore, the broken O($4$) symmetry is restored and the spectrum
corresponding to the $\Phi$ sector is organized as the O($4$)
multiplets. In addition, the corresponding excitations acquire an
energy gap which is given by
\begin{equation}
 \bar{\Delta}_s=c\Delta_ee^{-\pi \rho_0} \ , \label{gap1}
\end{equation}
where $c=O(1)$ is a non-universal constant. Because the charge
excitations in this phase are associated with the fermion fields
and $\bar{\Delta}_s\ll \Delta_e$, the mean-field gap $\Delta_e$
can be identified as the charge gap and the spin gap is determined
by $\bar{\Delta}_s$. Although the charge gap is not equivalent to
the spin gap, there is no spin-charge separation in this phase. In
conclusion, the long rage excitonic order becomes a short-ranged
one and all excitations are gapped.

A residual Coulomb interaction like the $g_3$ term in Eq.
(\ref{act2}) leads to the breaking of the O($4$) symmetry and the
mean-field theory favors $\bm{\Phi}_t\neq 0$. Following the
similar procedure, the Lagrangian which describes the
corresponding low-lying fluctuations now consists of ${\cal
L}_{\theta}$ as shown in Eq. (\ref{lag2}) and the O($3$)
non-linear $\sigma$ model. We have no arguments to exclude the
possibility of the appearance of the $\theta$-term in the O($3$)
non-linear $\sigma$ model as the case of the spin-$\frac{1}{2}$
Heisenberg chain. If a $\theta$-term exists, then the ground state
will exhibit an algebraic long range triplet excitonic (SDW) order
and the spin gap will vanish. On the other hand, if the
$\theta$-term vanishes as the two-leg spin ladder, then the spin
excitations still have a finite gap.

%%%%%%%%%%%%%%%%%%%%%%%%%%%%%%%%%%%%%%%%%%%%%%%%%%%%%%%%%%%%%%%%%%%%%%%%%%%%
\section{Conclusions and discussions}

In the present paper, we study the low energy physics of the
undoped nanographite ribbons with zigzag edges by neglecting the
electron-phonon interactions. We show that the interactions
between electrons substantially change the physics of the partly
flat bands. According to the above analysis, there are two
possible phases: a metallic phase with a spin gap (spin liquid)
and enhanced SS correlations (region II in Fig. \ref{pd}), and an
insulating phase with a gapped spectrum (region I in Fig.
\ref{pd}). Our results that there are two possible phases are
valid for both the cases of the single layer and
$AB$-stacking\cite{Khv}. In the latter case, the values of $v$,
$g$, $\tilde{g}$, and the exponent $m$ in Eq. (\ref{h1}) will be
different from those in the single layer. Through these parameters
$v$, $g$, $\tilde{g}$, and $m$, the material properties of the
graphite ribbon determine in which phase this system is truly
located.

To determine which one, the EI, the spin liquid, or the gapless
edge states predicted by the band theory, is the ground state of
the graphite ribbon with zigzag edges, we suggest two types of
experiments: (i) the measurement of the single-particle DOS by the
STM, and (ii) the measurement of the uniform magnetic
susceptibility. For the EI and spin liquid, the single-particle
DOS vanishes when the energy is smaller than the spectral gap.
This is distinguished from the gapless edge states predicted by
the band theory, where the single-particle DOS still has a finite
value at low energy. Another distinction between the SL and the
gapless edge states is the temperature dependence of the uniform
magnetic susceptibility. At low temperature, it will exhibit the
activated behavior in the SL due to the spin gap, whereas the
Curie-like behavior is expected for the gapless edge
states\cite{WFAS}.

The enhanced SS fluctuations arise from the singlet Cooper-pair
tunneling between two bands. It is also the origin of the spin
liquid phase with strong SS correlations in many one-dimensional
two-band models as emphasized in Ref. \onlinecite{EK}. In
contrast, the excitonic fluctuations are suppressed after taking
into account the quantum fluctuations. This results from a special
feature of the broken non-Abelian symmetry in $1+1$ dimensions,
where the corresponding non-linear sigma model becomes IR unstable
due to the lack of the topological term.

In the usual one-dimensional two-band electron system, the
competing order of the singlet superconductor is the inter-channel
CDW\cite{SMHG}, which also results from the same interaction as
the $\tilde{g}$ term in our Hamiltonian (\ref{h1}). The onset of
the inter-channel CDW requires the nearly equal electron density
in the two bands. This is impossible in our case because, here,
one band is empty and the other is completely filled.

Finally, we discuss the effects of electron or hole doping. For
finite doping, the system may be described by the one-band
Luttinger liquid with a very small Fermi velocity. For lightly
doping, the mean-field theory is supposed to be robust. Therefore,
our results are not affected qualitatively. According to the
analysis of Ref. \onlinecite{BVBG} on the doped EI, there may be
an inhomogeneous state between the EI and the Luttinger liquid
upon increasing the doping concentration.

%%%%%%%%%%%%%%%%%%%%%%%%%%%%%%%%%%%%%%%%%%%%%%%%%%%%%%%%%%%%%%%%%%%%%%%%%%%%
\acknowledgments

Y.L. Lee would like to thank M.F. Lin for discussions and Global
fiberoptics, Inc. for financial support. The work of Y.-W. Lee was
supported by the National Science Council of Taiwan under grant
NSC91-2112-M-029-012.


\begin{thebibliography}{99}
\bibitem{foot1} In $2+1$ or $3+1$ dimensions, $\langle \bm{J}_a \rangle \neq 0$
                is allowed. In that case, the term linear in $\partial_{\tau}\Phi$
                can exist as the FM case and the dispersion relation of the
                Goldstone modes at long wavelength will behave as: $\omega \sim
                |\vec{k}|^2$.
\end{thebibliography}
\end{document}